\documentclass[prl,twocolumn,groupedaddress,superscriptaddress,
amsfonts,amssymb,amsmath,floatfix,nofootinbib]{revtex4}

\usepackage{graphicx}
\usepackage{dcolumn}
\usepackage{bm}

\usepackage{array}
\begin{document}

%\preprint{}

\title{Maximum relative excitation of a specific vibrational mode \\
via optimum laser pulse duration}

% Please use the appropriate macro for each each type of information.

\author{Xiang Zhou}
%\thanks{}
\affiliation{Department of Physics, Wuhan University, Wuhan 430072,
China}
\affiliation{Department of Physics, Texas A\&M University, College
Station, Texas 77843, USA}
\author{Zhibin Lin}
\affiliation{Department of Physics, Texas A\&M University, College
Station, Texas 77843, USA}
\affiliation{Renewable Energy Materials Research Science and Engineering 
Center, and Department of Physics, Colorado School of Mines, Golden,
Colorado 80401, USA}
\author{Chenwei Jiang}
\affiliation{Department of Applied Physics, Xi'an Jiaotong University, 
Xi'an 710049, China}
\affiliation{Department of Physics, Texas A\&M University, College
Station, Texas 77843, USA}
\author{Meng Gao}
\affiliation{Department of Physics, Texas A\&M University, College
Station, Texas 77843, USA}
\author{Roland E. Allen}
\email[]{allen@tamu.edu}
\affiliation{Department of Physics, Texas A\&M University, College
Station, Texas 77843, USA}


\date{\today}

\begin{abstract}
For molecules and materials responding to femtosecond-scale optical 
laser pulses, we predict maximum relative excitation of a Raman-active 
vibrational mode with period $T$ when the pulse has an FWHM duration 
$\tau \approx 0.42\, T$. This result follows from a general analytical  
model, and is precisely confirmed by detailed 
density-functional-based dynamical simulations for C$_{60}$ and a carbon 
nanotube, which include anharmonicity, nonlinearity, no assumptions 
about the polarizability tensor, and no averaging over rapid 
oscillations within the pulse. The mode specificity is, of course, 
best at low temperature and for pulses that are electronically 
off-resonance, and the energy deposited in any mode is 
proportional to the fourth power of the electric field.
\end{abstract}

%\keywords{}

\maketitle

%\section{}
%\subsection{}
%\subsubsection{}
%\section{Introduction}

For a quarter century there has been considerable interest in
optimizing the vibrational response of molecules and materials to
ultrafast laser pulses~\cite{Nelson-1985-CPL,Nelson-1985-JCP,
Nelson-1990,Dresselhaus-1992,Shank-1995,
banin-1994,smith-1996,Nazarkin-1998,Nazarkin-1999,merlin-1999,
Kapteyn-2003,Mathies-2004,
Corkum-2004,Torralva2001,Zhang}. This problem is directly 
relevant to the broader issue of coherent control in physical, 
chemical~\cite{Rabitz-2000,Rabitz-Kapteyn-2001,Murnane-2002,
Gerber-control-1998}, and biological~\cite{Tsen-2007,Tsen-Sankey-2007,
Sankey-2008,Sankey-2009} systems.

Here we consider excitation via impulsive stimulated Raman scattering 
and related techniques using femtosecond-scale optical pulses. We find 
that the optimum full-width-at-half-maximum (FWHM) pulse 
duration for exciting a specific vibrational mode with angular 
frequency $2 \pi / T$ is given by $\tau \approx 0.42\, T$.
Our prediction results from a general analytical model, and is 
precisely confirmed by completely independent density-functional-based 
simulations for C$_{60}$~\cite{Dexheimer1993,Hohmann1994,Boyle2005,
Larrmann2007,Bhardwaj2003} and a small carbon 
nanotube~\cite{Dresselhaus-2008,Liu2002,Hulman2004}. Unlike the model,
these simulations include anharmonic effects in the vibrations,
nonlinear effects in the response to the applied field, and no
simplifying assumptions about the electronic polarizability tensor.

Our general model consists of the following: 

(1) The electric field has the form
\begin{equation}
\boldsymbol{\mathcal{E}}\left( t \right) = 
\boldsymbol{\mathcal{E}}_{0} \sin \left( \pi t/ 2 \tau \right) 
\sin \left( \omega t + \delta \right) \quad , \quad 0 < t < 2 \tau
\label{eq-1-2009}
\end{equation}
with $\omega \gg \pi / \tau $ and $\hbar \omega $ 
off resonance. Since the oscillations of 
$\sin ^{2} \left( \omega t + \delta \right) $
will average out to $1/2$ over a period that is 
short compared to the response time of the vibrating nuclei, we will
actually replace the square of Eq.~(\ref{eq-1-2009}) 
by the envelope function
\begin{equation}
\bar{\mathcal{E}}\left( t \right)^{2} = 
\bar{\mathcal{E}}_{0}^{2}
\sin ^{2} \left( \pi t/ 2 \tau \right) \quad , \quad 0 < t < 2 \tau 
\label{eq-2-2009}
\end{equation}
with $\bar{\mathcal{E}}_{0}^{2} = \mathcal{E}_{0}^{2}/2$.
This form has the following nice features~\cite{Graves}: (i)~The duration 
is finite and need not be truncated. (ii)~The FWHM duration is exactly 
half the full duration $2 \tau $. (iii)~A plot reveals that it 
closely resembles a Gaussian. (iv)~The slope is zero at beginning and end. 

(2) The initial conditions 
$Q_{k} \left( 0 \right) = \dot{Q}_{k} \left( 0 \right) = 0$ 
are imposed on the 
normal-mode coordinates $Q_{k} \left( t \right) $. This approximation is 
valid for the expectation value below Eq.~(\ref{eq-6-2009}) at low temperature, 
or an ensemble average at higher temperature, in the linearized 
Eq.~(\ref{eq-3-2009}).

(3) The equation of motion is given by the standard (Placzek) model 
for Raman-active modes~\cite{Bloembergen,Kapteyn-2003,Tsen-Sankey-2007}:
\begin{equation}
d^{2}{Q}_{k}/dt^{2}+\omega _{k}^{2}Q_{k}=\alpha _{k}^{\prime }
\mathcal{E} \left( t \right)^{2}/2  
\label{eq-3-2009}
\end{equation}
where the anharmonic and damping terms (in the normal-mode coordinates 
$Q_{k}$), nonlinear terms (in the electric field $\boldsymbol{\mathcal{E}} 
\left( t\right) $), and off-diagonal terms (in the polarizability 
tensor $\boldsymbol{\mathcal{ \alpha }}$) have been neglected, with 
the electric dipole moment given by 
$\boldsymbol{\mu}_{\mathrm{tot}} = \boldsymbol{\mu} + 
\boldsymbol{\mu} _{\mathrm{ind}}$,
$\boldsymbol{\mu} _{\mathrm{ind}} = 
\boldsymbol{\alpha} \cdot \boldsymbol{\mathcal{E}} + \ldots$,
$\alpha _{xx} = \alpha _{0} + \sum_{k}\alpha _{k}^{\prime } Q_{k } 
+ \ldots$, and the radiation field taken to be linearly polarized 
in the $x$-direction. Equation (\ref{eq-3-2009}) follows from e.g. 
the Heisenberg equations of motion for the quantized Hamiltonian 
\begin{equation}
\hat{H}=\sum_{k}\left( \hat{P}_{k}^{2}/2 + 
\omega _{k}^{2}\hat{Q}_{k}^{2}/2 \right) - \hat{\boldsymbol{\mu}}
\cdot \boldsymbol{\mathcal{E}} - \boldsymbol{\mathcal{E}} \cdot 
\hat{\boldsymbol{\alpha}}\cdot \boldsymbol{\mathcal{E}}/2
\label{eq-6-2009}
\end{equation}
with $Q_{k}\equiv \left \langle \hat{Q}_{k}\right \rangle $ and 
$\boldsymbol{\alpha} _{k}^{\prime }\equiv \left \langle 
\partial \hat{\boldsymbol{\alpha}}/\partial \hat{Q}_{k}\right 
\rangle $, when the term involving 
$\partialÊ\hat{\boldsymbol{\mu}} / \partial Q_{k}Ê$ 
vanishes or is neglected (e.g. in 
taking rapid oscillations of the electric field 
$\boldsymbol{\mathcal{E}}\left( t \right)$ to average out to zero, an 
approximation which also causes the lowest-order nonlinear term to 
vanish in Eq.~(\ref{eq-6-2009})).

If Eq.~(\ref {eq-2-2009}) is substituted into Eq.~(\ref{eq-3-2009}), 
the solution after the pulse is found to reduce to
\begin{eqnarray}
\hspace {-0.5cm} Q_{k}(t) &=& \frac{\alpha _{k}^{\prime } 
\mathcal{E}_{0}^{2}}
{4 \omega _{k}^{2}}\frac{\sin (r_{k}\pi )}{1 - r_{k}^{2}}
\sin (\omega _{k}t-r_{k}\pi ) \quad , \quad t \geq 2 \tau 
\label{eq-7-2009}
\end{eqnarray}
where
\begin{equation}
r_{k} = 2 \tau /T_{k} \; .
\label{eq-7a-2009}
\end{equation}
The total energy 
$\left( \dot{Q}_{k}^{2}+\omega _{k}^{2}Q_{k}^{2}\right) /2$ in
vibrational mode $k$ is therefore
\begin{equation}
E_{k}=\frac{1}{32} \left[ \frac{\alpha _{k}^{\prime}}
{\omega _{k}} \frac{\sin \left( r_{k}\pi \right)}{1 - r_{k}^{2}}\right]
^{2}\mathcal{E}_{0}^{4}  \quad , \quad t \geq 2 \tau \; .
\label{eq-8-2009}
\end{equation}
$E_{k}$ is, of course, equivalent to the maximum kinetic energy 
$K_{k}^{\mathrm{max}}$. 

The maximum response is then given by the extremum of the function in
brackets, which occurs at $r_{k}\approx 0.8375$, or $2 \tau \approx
0.8375 \, T_{k}$.  

We have tested this prediction by performing independent supercomputer 
simulations for C$_{60}$ and a small carbon nanotube, using the
density-functional-based approach of Frauenheim and 
co-workers~\cite{Porezag1995,Seifert1996}, together with  
semiclassical electron-radiation-ion dynamics (SERID), which is 
defined by the following equations~\cite{allen-2008}:

(1) Time-dependent Schr\"{o}dinger equation in a nonorthogonal basis:
\begin{equation}
i\hbar \, \partial \boldsymbol{\psi} \left(  t \right)/\partial t 
=\boldsymbol{S}^{-1}\cdot\boldsymbol{H}\cdot
\boldsymbol{\psi}\left(  t\right) \; .
\label{eq-9-2009}
\end{equation}
With $60$ atoms and a minimal basis set, the matrices are $240 \times 
240$. A time step of 50 attoseconds was used. The simulation time 
after completion of the pulse was 2000 fs for C$_{60}$ and 1000 fs 
for the nanotube. 

(2) Ehrenfest's theorem (in a nonorthogonal basis):  
\begin{eqnarray}
M \frac{d^{2} X}{dt^{2}} = 
-\frac{1}{2}\sum_{n} \boldsymbol{\psi }_{n}^{\dagger } 
\cdot \left( \frac{\partial \boldsymbol{H}}{\partial X } 
 - i\hbar \frac{\partial \boldsymbol{S}}{\partial X}
\frac{\partial }{\partial t}\right) \cdot \boldsymbol{\psi}_{n} \nonumber \\
 + h.c. - \frac{\partial U_{rep}}{\partial X} 
 \label{eq-10-2009}
\end{eqnarray}
where $X$ is any nuclear coordinate. 
The Hamiltonian matrix $\boldsymbol{H}$, overlap matrix
$\boldsymbol{S}$, and effective ion-ion repulsion $U_{rep}$ were  
determined by the methods and results of Refs. \cite{Porezag1995}   
and \cite{Seifert1996} and later work by this group. 

(3) Coupling of electrons to the radiation field through the
time-dependent Peierls substitution 
\begin{eqnarray}
H \left(  \ell^{\prime},\ell\right) 
=H_{0}\left( \ell^{\prime},\ell \right)  e^{iq \boldsymbol{A} 
\left( t \right) 
\cdot\left( \boldsymbol{X}^{\prime}-\boldsymbol{X}\right)  /\hbar c}
\label{eq-11-2009}
\end{eqnarray}
where $q=-e$. $\boldsymbol{A}$ is the vector potential, which 
in the present simulations was taken to have the form 
\begin{equation}
\boldsymbol{A} \left( t \right) = \boldsymbol{A}_{0}\sin 
\left( \pi t/ 2 \tau \right) 
\cos (\omega t) \quad ,\quad 0 \leq t \leq 2 \tau 
\label{eq-12-2009}
\end{equation}
with $\omega \gg \pi / \tau $, so that  $\boldsymbol{\mathcal{E}}(t) 
\approx \omega \boldsymbol{A}_{0} \sin \left( \pi t/ 2 \tau \right)
\sin (\omega t)$, although Eq.~(\ref{eq-12-2009}) was actually used. 
The polarization vector was taken to lie along the $x$-axis, with 
the $z$-axis pointing down the axis of a nanotube. 

Within the present DFT-based model, the energy
gap for electronic excitations is $1.80$ eV for C$_{60}$ and $1.56$ eV for 
a $(3,3)$ nanotube with a periodicity length of 5 unit cells. (This model 
nanotube, with only $60$ atoms, has a substantial gap because small 
wavenumbers $k_{z}$ are not allowed. An infinitely long (3,3) nanotube
would be metallic~\cite{Liu2002}.) The laser pulse photon energy was 
chosen to be $0.69$~eV for C$_{60}$ and $0.80$~eV for the nanotube, 
and is thus off-resonance in both cases. 

For C$_{60}$, seven of the ten Raman-active modes were appreciably
excited: modes $A_g(1)$, $A_g(2)$, $H_g(1)$, $H_g(4)$, $H_g(5)$,
$H_g(6)$ and $H_g(8)$, with periods of $59.7$ fs, $20.6$ fs, $125$ fs,
$38.5$ fs, $27.0$ fs, $23.0$ fs, and $19.0$ fs respectively.
For the $(3,3)$ nanotube only an $E_g$ mode with a period of $182$ fs 
was observed, and this is consistent with the experimental difficulty 
of observing the radial breathing mode in the Raman 
spectrum~\cite{Hulman2004}. 

\begin{figure}[hb]
 \includegraphics[width=3.4in]{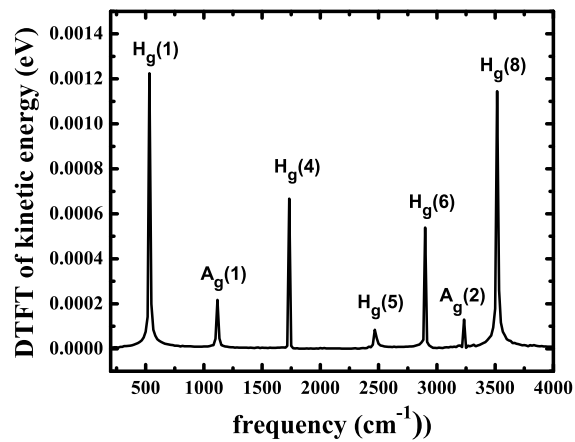}
 \caption{Numerical Fourier transform of the total kinetic energy of 
C$_{60}$ after being subjected to a laser pulse with a FWHM duration 
$\tau $ of $12$ fs, photon energy $\hbar \omega $ of $0.69$ eV, 
and electric field amplitude $\mathcal{E}_{0}$ of $5.0$ V/nm.
Seven Raman-active modes are clearly observed, each with a frequency 
$\omega _{k}/ 2 \pi$ which is half the frequency exhibited in the figure 
by the corresponding kinetic energy $K_{k}$.
\label{Fourier}}
 \end{figure}

As in Ref.~\cite{Zhang}, it is natural to
characterize the strength of the vibrational response of a specific 
mode by its maximum kinetic energy $K_{k}^\mathrm{max}$.
In the harmonic approximation, after completion of the laser pulse, the 
velocity $\dot{Q}_{k}(t)$ is proportional to 
$\cos \left( \omega_{k} t + \delta \right) $, so $K_{k}$ is proportional to 
$\cos^2 \left( \omega_{k} t + \delta \right) = 
\left[ 1 + \cos \left( 2 \omega_{k} t + 2 \delta \right) \right] / 2 $. 
A numerical Fourier transform of the total kinetic energy therefore
shows a peak at $2 \, \omega_{k}$, with a strength proportional to the
response of the normal mode with angular frequency $\omega_{k}$. 
Figure~\ref{Fourier} shows our results for the Fourier
transform of the total kinetic energy for C$_{60}$ following a 
$12$ fs, $0.69$ eV, $5.0$ V/nm pulse.

\begin{figure}[ht]
 \includegraphics[width=3.4in]{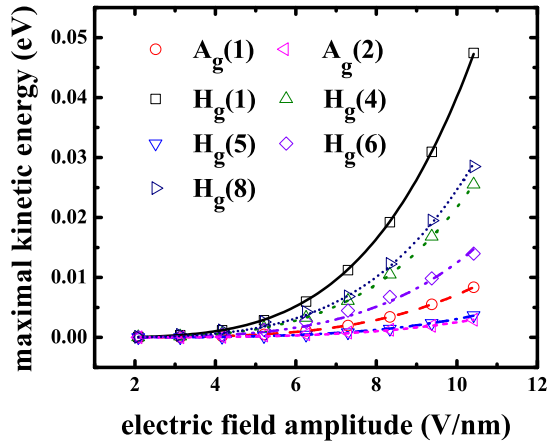}
 \caption{Maximum kinetic energy of $A_g(1)$, $A_g(2)$, $H_g(1)$,
$H_g(4)$, $H_g(5)$, $H_g(6)$ and $H_g(8)$ modes of C$_{60}$ 
responding to  $0.69$ eV laser pulses when the FWHM pulse duration 
$\tau $ is fixed at $12$ fs and the electric field amplitude 
$\mathcal{E}_{0}$ is varied. Here, and in the subsequent figures, 
each point represents a DFT-based SERID simulation and 
each curve has the form predicted by Eq.~(\ref{eq-8-2009}), with no
adjustable parameters except the effective polarizability parameter
$\alpha _{k}^{\prime } $, which is determined by a least-squares fit. 
\label{C60intensity}}
 \end{figure}

Figure~\ref{C60intensity} shows that the detailed simulations agree with
Eq.~(\ref{eq-8-2009}) regarding the dependence of the maximum kinetic 
energy $K_{k}^{\mathrm{max}}$ of mode $k$ on the field amplitude 
$\mathcal{E}_{0}$ (for the range of amplitudes considered here) when
the pulse duration is fixed: 
$K_{k}^{\mathrm{max}} \propto \mathcal{E}_{0}^{4}$.
 
\begin{figure}[hb]
 \includegraphics[width=3.4in]{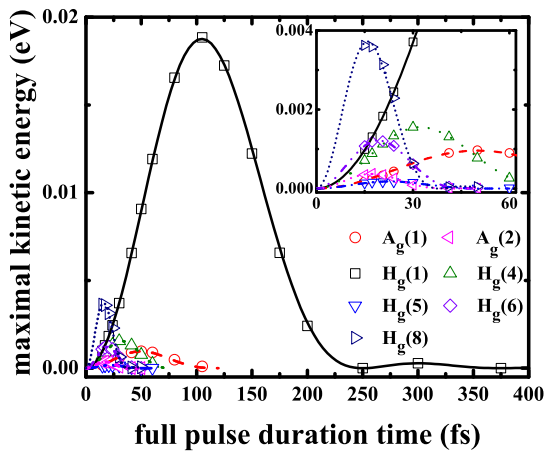}
 \caption{Maximum kinetic energy for the 7 prominent modes of C$_{60}$
responding to  $0.69$ eV laser pulses when 
$\mathcal{E}_{0}$ is fixed at $5.0$ V/nm and the full pulse duration
time $2 \tau $ is varied. The agreement between the general model of 
Eqs.~(\ref{eq-2-2009})-(\ref{eq-3-2009}) (curves) and the 
density-functional-based simulations of 
Eqs.~(\ref{eq-9-2009})-(\ref{eq-12-2009}) (points) is remarkable.
\label{C60duration}}
 \end{figure}

When $\mathcal{E}_{0}$ is fixed, on the other hand, 
$K_{k}^{\mathrm{max}}$ is predicted to be proportional to the square of 
the function involving $r_{k}$ in Eq.~(\ref{eq-8-2009}), and is maximal 
when $r_{k} \approx 0.84$. Figure~\ref{C60duration} shows, for example, 
that the $H_g(8)$ mode in C$_{60}$ dominates for $2 \tau < 23.4$ fs, 
and the $H_g(1)$ mode for $2 \tau > 23.4$ fs, in agreement 
with the analytical model of Eqs.~(\ref{eq-2-2009})-(\ref{eq-3-2009}), 
and in qualitative agreement with 
other simulations~\cite{Zhang} and experiment
~\cite{Dexheimer1993,Hohmann1994,Bhardwaj2003,Boyle2005,Larrmann2007}.
 
In Fig.~\ref{C60duration}, the agreement between the predictions of 
the analytical model (curves) and the completely independent DFT-based 
simulations (points) is truly remarkable. The only adjustable
parameter for each curve is the effective polarizability parameter
$\alpha _{k}^{\prime } $.

\begin{figure}[hb]
 \includegraphics[width=3.4in]{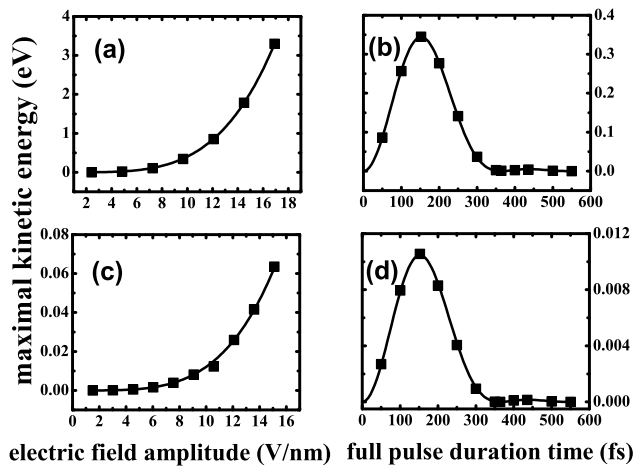}
 \caption{Maximum kinetic energy of $E_g$ mode for (3,3) carbon 
nanotube, excited by a $0.8$ eV laser pulse (a) when $\tau $ is
fixed at $76$ fs and $\mathcal{E}_{0}$ is varied and (b) when
$\mathcal{E}_{0}$ is fixed at $9.66$ V/nm and $\tau $ is varied.
Panels (c) and (d) are the corresponding results for a pulse with a
photon energy of $2.0$ eV. 
\label{CNT33}}
 \end{figure}

There is similarly remarkable agreement for the carbon nanotube, as
can be seen in Fig.~\ref{CNT33}. In this case, two photon energies
$\hbar \omega $ were used: $0.8$ eV and $2.0$ eV, which are
respectively below and above the excitation gap of $1.56$ eV for 
this model nanotube, as discussed above. 
Figure~\ref{CNT33} shows the results for 
the maximum kinetic energy $K_{k}^{\mathrm{max}}$ of the one prominent 
mode observed in this case. (Recall that the polarization vector for
the pulse was chosen to point across the nanotube, so that only
radial modes should be directly excited, and this is a severe
constraint for the small (3,3) nanotube, as is also observed in
experiment~\cite{Hulman2004}.) The panels in Fig.~\ref{CNT33}
correspond to: (a) $\hbar \omega = 0.8$ eV, $\tau = 76$ fs; 
(b) $\hbar \omega = 0.8$ eV, $\mathcal{E}_{0}= 9.66$ V/nm;
(c) $\hbar \omega = 2.0$ eV, $\tau = 76$ fs; 
(d) $\hbar \omega = 2.0$ eV, $\mathcal{E}_{0}= 9.66$ V/nm.
It is interesting that the analytical model still works in this case even 
for above-the-gap excitation, but we find less satisfactory agreement 
for C$_{60}$, and certainly do not expect the model of 
Eq.~(\ref{eq-3-2009}) to be generally valid when there is a substantial 
population of electrons in excited states at the end of the pulse. In 
this case other effects will ordinarily be of dominant 
importance~\cite{Dresselhaus-1992,Torralva2001}.
 
As mentioned above, the electronic response in the DFT-based simulations 
can be characterized by an effective polarizability parameter
$\alpha _{k}^{\prime } $ if one fits Eq.~(\ref{eq-8-2009}) to the results 
of the simulations. If the model leading to Eq.~(\ref{eq-8-2009}) 
is in fact consistent with the detailed simulations, then the values 
obtained when (a) varying the electric 
field amplitude $\mathcal{E}_0$ at constant 
(FWHM) pulse duration $\tau $ and (b) varying $\tau $ at constant 
$\mathcal{E}_0$ should also be reasonably 
consistent. Table \ref{tab} shows the results for all of the prominent 
Raman-active modes.  
The two procedures do lead to generally consistent values, with 
the differences for the $A_g(2)$, $H_g(6)$, and $H_g(8)$ modes
presumably arising from their short periods, so that the approximation
of averaging out the $\omega$ oscillations in Eq.~(\ref{eq-2-2009}) is
not valid. However, the prediction 
regarding optimum pulse duration works even for these modes, as can be
seen in Fig.~\ref{C60duration}.
%%%%%%%%%%%%%%%%%%

% tables should appear as floats within the text
% The ruledtabular enviroment adds doubled rules to table and sets a
% reasonable default table settings.
% Use the table* environment to get a full-width table in two-column
% Add \usepackage{longtable} and the longtable (or longtable*}
% environment for nicely formatted long tables. Or use the the [H]
% placement option to break a long table (with less control than 
 \begin{table}%[H] add [H] placement to break table across pages
 \caption{\label{tab} For each of the prominent modes observed in the
response of C$_{60}$ and the (3,3) carbon nanotube, the values of the 
effective $\alpha _{k}^{\prime } $ were obtained through the
procedures designated (a) and (b) in the text. In procedure (a), 
$\mathcal{E}_0$ was varied while the (FWHM) duration $\tau $ was fixed 
at $12$ fs, for C$_{60}$, or $76$ fs, for the nanotube. 
In procedure (b), $\tau $ was varied while $\mathcal{E}_0$ was fixed
at $5.0$ V/nm, for C$_{60}$, or $9.66$ V/nm, for the nanotube.
The nanotube was also subjected 
to $2.0$ eV pulses, and the effective values of 
$\alpha _{k}^{\prime } $ obtained through both procedures are again
shown.}
 \begin{ruledtabular}
\begin{tabular}{cccc}
& mode & $\alpha _{k}^{\prime } $ (a)
& $\alpha _{k}^{\prime } $ (b) \\
carbon nanotube & E$_g$ æ& $7.578\times10^{-4}$ & $7.496\times10^{-4}$ \\
C$_{60}$ æ& A$_{g}(1)$ æ& $4.416 \times 10^{-4}$ & $4.505 \times ÊÊÊÊÊÊÊÊÊÊÊÊÊÊÊ
10^{-4}$\\
& A$_g$(2) & $6.248 \times 10^{-4}$ & $8.102\times 10^{-4}$\\
& H$_g$(1) æ& $9.668 \times 10^{-4}$ & $9.516 \times 10^{-4}$\\
& H$_g$(4) æ& $8.978 \times 10^{-4}$ & $8.884 \times 10^{-4}$\\
& H$_g$(5) æ& $4.482 \times 10^{-4}$ & $4.294 \times 10^{-4}$\\
& H$_g$(6) æ& $1.127 \times 10^{-3}$ & $1.316 \times 10^{-3}$\\
& H$_g$(8) æ& $2.376 \times 10^{-3}$ & $2.788 \times 10^{-3}$\\
\hline
nanotube, above gap $\hbar \omega $
& E$_g$ æ& $1.314 \times 10^{-4}$ & $1.307 \times 10^{-4}$ \\
% Lines of table here ending with \\ ÊÊÊÊÊÊÊÊÊÊÊÊÊÊÊÊÊÊÊÊÊÊÊÊÊÊÊÊÊÊÊÊÊÊÊÊÊÊÊÊÊÊÊ
\end{tabular}
\end{ruledtabular}
 \end{table}

It has long been suspected that there might be an optimum pulse 
duration. For example, in Ref.~\cite{Zhang} the pulse
envelope was assumed to have a Gaussian form:
$\left| \mathcal{E}(t) \right|^{2} = A \exp \left( - 2 t^{2}/\bar{\tau}^{2} 
\right) \cos ^{2} \left( \omega t \right) $
(with a slight change in notation). Simulations were then performed 
with a model that essentially lies between the analytical model of 
Eqs.~(\ref{eq-2-2009})-(\ref{eq-3-2009}) and our much more extensive
density-functional-based 
simulations, defined by Eqs.~(\ref{eq-9-2009})-(\ref{eq-12-2009}).
It was found that the optimum value of $\bar{\tau}$ was about $T / 3.4$. 
The corresponding FWHM duration then has an optimum value 
$\tau = \sqrt{2 \ln 2 } \; \bar{\tau} \approx 0.35 \, T$. 
In both our general model and DFT-based simulations, on the other hand, 
the optimal FWHM pulse duration is found to be given by
$\tau \approx 0.42 \, T$. Both results seem to be consistent with the
existing experiments~\cite{Dexheimer1993,Hohmann1994,Bhardwaj2003,
Boyle2005,Larrmann2007}, and both are larger than the na\"{i}ve prediction
of $\tau = 0.25 \, T$. In our result, the reason appears to be the
phase lag apparent in Eq. (\ref{eq-7-2009}).

It should be emphasized that our result of $\tau \approx 0.42 \, T$ 
is for the maximum response of this mode when the
intensity of the laser pulse is kept fixed while the duration 
is varied. On the other hand, if the fluence is instead held fixed
(while  $\tau$ is varied), Eqs.~(\ref{eq-7-2009}) and (\ref{eq-8-2009}) 
are changed to 
\begin{equation}
Q_{k}(t)=\frac{\alpha _{k}^{\prime}W}{\omega_{k}\pi c\epsilon_0}
\frac{\sin (r_{k} \pi )}{r_k (1 - r_{k}^{2})}\sin (\omega _{k}t-r_{k}\pi ) 
\quad , \quad t \geq 2 \tau
\label{eq-11a-2009}
\end{equation}
and
\begin{equation}
E_{k}=\frac{1}{2} \left[ \frac{\alpha_{k}^{\prime}W}{\pi c\epsilon_0} 
\frac{\sin \left( r_{k}\pi \right)}{r_k(1- r_{k}^{2})}\right]^{2} Ê
\quad , \quad t \geq 2 \tau 
\label{eq-12a-2009}
\end{equation}
where
\begin{equation}
W=\frac{c\epsilon_0}{4}\mathcal{E}_0^2\tau 
\label{eq-13-2009}
\end{equation}
is the energy in the pulse. If $W$ is held constant, the maximum in 
Eq.~(\ref{eq-12a-2009}) is achieved as $r_{k} \rightarrow 0$:
\begin{equation}
E_{k} \left( 0 \right) =\frac{1}{2} 
\left[ \frac{\alpha_{k}^{\prime}W}{c\epsilon_0}\right]^{2}
\label{eq-14-2009}
\end{equation}
so that
\begin{equation}
\frac{E_{k} \left( \tau \right)}{E_{k} \left( 0 \right)} =
\left[\frac{\sin \left( r_{k}\pi \right)}{\pi r_k(1- r_{k}^{2})}\right]^{2}
\; .
\label{eq-15-2009}
\end{equation}
Then $r_{k} = 0.84$ gives
$E_{k} \left( 0.42 \, T \right) / E_{k} \left( 0 \right) = 0.385$.
However, Eq. (\ref{eq-14-2009}) holds for all modes, so there is no
\textit{relative} enhancement of any preferred mode.

It should also be emphasized that the success of the simple analytical
model of Eq. (\ref{eq-3-2009}) results from the fact that the additional
effects described below Eq. (\ref{eq-3-2009}) are not very important for
the pulse intensities, time durations, and systems considered here.
The various effects omitted in the model will lead to richer (or
messier) behavior in regimes where nonlinear effects, anharmonicity, 
damping, etc. become comparable in importance to the lowest-order
effects included in the model. In future studies with 
density-functional-based simulations based on 
Eqs.~(\ref{eq-9-2009})-(\ref{eq-12-2009}) it may be interesting
to sort out the influence of the higher-order effects.

Finally, as emphasized below Eq. (\ref{eq-1-2009}), the present results are 
for $\hbar \omega $ off resonance, in contrast to those of e.g. 
Refs. \cite{Torralva2001} and \cite{smith-1996}. In particular, Smith 
and Cina considered pulses whose central frequencies are near 
resonance with an electronic transition, in a model with  
two electronic levels and one vibrational degree of freedom. They
found that the momentum increment (for the nuclei) falls off less 
rapidly with increasing offset (from resonance) than does the
population loss (from the electronic ground state), and they 
presented a rather complex strategy for optimizing the various 
parameters for these ``preresonant'' pulses. The work of this group
and others is thus complementary to that of the present paper.

In summary, we find remarkable agreement between the general model 
of Eqs.~(\ref{eq-2-2009})-(\ref{eq-3-2009}) and the detailed DFT-based 
simulations based on Eqs.~(\ref{eq-9-2009})-(\ref{eq-12-2009}) for 
C$_{60}$ and a carbon nanotube. At fixed pulse intensity, 
both of these approaches predict maximum 
excitation of a Raman-active vibrational mode with period $T$ when the 
pulse has an FWHM duration $\tau \approx 0.42\, T$. 

% if in two-column mode, this environment will change to single-column
% format so that long equations can be displayed. Use
%\begin{widetext}
%\end{widetext}

% Surround table environment with turnpage environment for landscape
% table
% \begin{table}
% \begin{ruledtabular}
% \end{tabular}
% \end{ruledtabular}
% \end{table}

%\appendix
%\section{}

\begin{acknowledgments}
This work was supported by the Robert A. Welch Foundation (Grant
A-0929) and the China Scholarship Council, and we wish to
thank the Texas A\&M University Supercomputing Facility for the use
of its parallel computing resources.
\end{acknowledgments}


\begin{thebibliography}{}
   
% vibrational control
\bibitem{Nelson-1985-CPL}
S. De Silvestri, J. G. Fujimoto, E. P. Ippen, E. B. Gamble Jr., 
L. R. Williams, and K. A. Nelson, Chem. Phys. Lett. \textbf{116}, 
146 (1985).
\bibitem{Nelson-1985-JCP}
Y.-X. Yan, E. B. Gamble Jr., and K. A. Nelson, J. Chem. Phys. 
\textbf{83}, 5391 (1985).
\bibitem{Nelson-1990}
A. M. Weiner, D. E. Leaird, G. P. Wiederrecht, and K. A. Nelson,
Science  \textbf{247}, 1317 (1990).
\bibitem{Dresselhaus-1992}
H. J. Zeiger, J. Vidal, T. K. Cheng, E. P. Ippen, G. Dresselhaus, and M. S.
Dresselhaus, Phys. Rev. B  \textbf{45}, 768 (1992).
\bibitem{Shank-1995}
C. J. Bardeen, Q. Wang, and C. V. Shank, Phys. Rev. Lett.  \textbf{75}, 
3410 (1995).
% more vibrational control
\bibitem{Nazarkin-1998}
A. Nazarkin and G. Korn, Phys. Rev. A \textbf{58}, R61 (1998).
\bibitem{Nazarkin-1999}
A. Nazarkin, G. Korn, M. Wittmann, and T. Elsaesser, Phys. Rev. Lett.
\textbf{83}, 2560 (1999).
\bibitem{Kapteyn-2003}
R. A. Bartels, S. Backus, M. M. Murnane, and H. C. Kapteyn, Chemical 
Physics Letters \textbf{374}, 326 (2003).
\bibitem{Mathies-2004}
S.-Y. Lee, D. Zhang, D. W. McCamant, P. Kukura, and R. A. Mathies, 
J. Chem. Phys.  \textbf{121}, 3632 (2004).
\bibitem{Corkum-2004}
H. Niikura, D. M. Villeneuve, and P. B. Corkum, Phys. Rev. Lett.
 \textbf{92}, 133002 (2004).
\bibitem{Torralva2001}
B. Torralva, T. A. Niehaus, M. Elstner, S. Suhai, Th. Frauenheim, 
and R. E. Allen, Phys. Rev. B  \textbf{64}, 153105 (2001).
\bibitem{Zhang}
G. P. Zhang and T. F. George, Phys. Rev. Lett.  \textbf{93}, 
147401 (2004); Phys. Rev. B \textbf{73}, 035422 (2006).
% requested by Referee 4
\bibitem{banin-1994}
U. Banin, A. Bartana, S. Ruhman, and R. Kosloff, J. Chem. Phys. 
\textbf{101}, 8461 (1994).
\bibitem{smith-1996}
T. J. Smith and J. A. Cina, J. Chem. Phys. \textbf{104}, 1272 (1996).
\bibitem{merlin-1999}
T. E. Stevens, J. Hebling, J. Kuhl, and R. Merlin, 
Physica Status Solidi (b) \textbf{215}, 81 (1999).
% coherent control
\bibitem{Rabitz-2000}
H. Rabitz, R. de Vivie-Riedle, M. Motzkus, and K. Kompa,
Science \textbf{288}, 824 (2000).
\bibitem{Rabitz-Kapteyn-2001}
T. C. Weinacht, R. Bartels, S. Backus, P. H. Bucksbaum, 
B. Pearson, J. M. Geremia, H. Rabitz, H. C. Kapteyn, and M. M. Murnane, 
Chem. Phys. Lett. \textbf{344}, 333 (2001).
\bibitem{Murnane-2002}
R. A. Bartels, T. C. Weinacht, S. R. Leone, H. C. Kapteyn, and M.
M. Murnane, Phys. Rev. Lett. \textbf{88}, 033001(2002).
\bibitem{Gerber-control-1998}
A. Assion, T. Baumert, M. Bergt, T. Brixner, B. Kiefer, 
V. Seyfried, M. Strehle, and G. Gerber, Science \textbf{282}, 
919 (1998).
% biological
\bibitem{Tsen-2007}
K. T. Tsen, S.-W. D. Tsen, C.-L. Chang, C.-F. Hung, T. C. Wu, 
and J. G. Kiang, Virol. J. \textbf{4}, 50 (2007).
\bibitem{Tsen-Sankey-2007}
K. T. Tsen, S.-W. D. Tsen, O. F. Sankey, and J. G. Kiang, 
J. Phys.: Condens. Matter \textbf{19}, 472201 (2007).
\bibitem{Sankey-2008}
E. C. Dykeman and O. F Sankey, Phys. Rev. Lett. \textbf{100}, 
028101 (2008).
\bibitem{Sankey-2009}
E. C. Dykeman and O. F Sankey, J. Phys.: Condens. Matter 
\textbf{21}, 505102 (2009). 
% C60
\bibitem{Dexheimer1993}
S. L. Dexheimer, D. M. Mittleman, R. W. Schoenlein, W. Vareka, X.
-D. Xiang, A. Zettl, and C. V. Shank, in \textit{Ultrafast Phenomena VIII},
edited by J. L. Martin, A. Migus, G. A. Mourou, and A. H. Zewail,
(Springer-Verlag, Berlin, 1993) p. 81.
\bibitem{Hohmann1994}
H. Hohmann, C. Callegari, S. Furrer, D. Grosenick, 
E. E. B. Campbell, and I. V. Hertel , Phys. Rev. Lett. \textbf{73}, 
1919 (1994).
\bibitem{Boyle2005}
M. Boyle, T. Laarmann, K. Hoffmann, M. Hed\'{e}n, E. E. B. Campbell, 
C. P. Schulz, and I. V. Hertel, Eur. Phys. J. D \textbf{36}, 339 (2005).
\bibitem{Larrmann2007}
T. Laarmann, I. Shchatsinin, A. Stalmashonak, M. Boyle, 
N. Zhavoronkov, J. Handt, R. Schmidt, C. P. Schulz, and I. V. Hertel, 
Phys. Rev. Lett. \textbf{98}, 058302 (2007).
% carbon nanotubes
\bibitem{Bhardwaj2003}
V. R. Bhardwaj, P. B. Corkum, and D. M. Rayner, Phys. Rev. Lett.
\textbf{91}, 203004 (2003).
\bibitem{Dresselhaus-2008}
\textit{Carbon Nanotubes, Advanced Topics in the Synthesis, Structure,
Properties and Applications}, edited by A. Jorio, G. Dresselhaus,
and M. S. Dresselhaus (Springer, Berlin, 2008).
\bibitem{Liu2002}
H. J. Liu and C. T. Chan, Phys. Rev. B \textbf{66}, 115416 (2002).
\bibitem{Hulman2004}
M. Hulman, H. Kuzmany, O. Dubay, G. Kresse, L. Li,
Z. K. Tang, P. Knoll, and R. Kaindl, Carbon \textbf{42}, 1071 (2004).
% analytical model
\bibitem{Graves}
J. S. Graves and R. E. Allen, Phys. Rev. B \textbf{58}, 13627 (1998).
\bibitem{Bloembergen}
Y. R. Shen and N. Bloembergen, Phys. Rev. \textbf{137}, A1787 (1965).
% SERID
\bibitem{Porezag1995}
D. Porezag, Th. Frauenheim, Th. K\"{o}hler, G. Seifert, and R.
Kaschner, Phys. Rev. B \textbf{51}, 12947 (1995).
\bibitem{Seifert1996}
G. Seifert, D. Porezag, and Th. Frauenheim Int. J. Quantum Chem.
\textbf{58}, 185 (1996). See http://www.dftb.org and
http://www.dftb-pluse.info for the currently best versions.
\bibitem{allen-2008}
R. E. Allen, Phys. Rev. B \textbf{78}, 064305 (2008) and references therein.

\end{thebibliography}

% figures should be put into the text as floats.
% Use the graphics or graphicx packages (distributed with LaTeX2e)
% and the \includegraphics macro defined in those packages.
% See the LaTeX Graphics Companion by Michel Goosens, Sebastian Rahtz,
% entire page. There is no need to do explicit centering.

% Surround figure environment with turnpage environment for landscape
% figure
% \begin{figure}
% \end{figure}

\end{document}